\documentstyle[12pt,epsfig]{article}
\def \ins#1#2#3#4#5#6 {
  \begin{figure}[#1]
    \begin{center}
      \psfig{file=#2,width=#3,height=#4,angle=0}
      \caption{#5}
      \label{#6}
    \end{center}
  \end{figure}
  }
\begin{document}
%\large

\begin{center}
{\Large \bf
% Version of UrQMD Model Code Installed at PANDA Framework\\
Patches to UrQMD Model Code
}
\end{center}

\begin{center}
A. Galoyan,
J. Ritman\footnote{University of Bochum and Research Center Julich,
                   Institute for Nuclear Physics (IKP),
                   Forschungszentrum Julich, Germany},
V. Uzhinsky\\
Joint Institute for Nuclear Research,\\
Dubna, Russia
\end{center}

\begin{center}
\begin{minipage}{12cm}
Patches to the code of the Ultra-relativisic Quantum Molecular Dynamics model (UrQMD
version 1.3) are described. They improve the code operation. The most important one
is done for the file ANNDEC.F.
%The improved code is installed at PANDA framework for simulation
%of anti-protion interactions with protons and nuclei.
\end{minipage}
\end{center}

\vspace{0.5cm}

%\section*{Introduction}

Ultra-relativistic Quantum Molecular Dynamics Model (UrQMD \cite{UrQMD1,UrQMD2}) is one of
the models applied for analysis of high energy hadron-nucleus and nucleus-nucleus interactions
at present time. Quite clear physical ideas put in its ground and their good program implementation
lead to its large popularity in high energy physics. There are a lot of various applications of
the model (see for example \cite{home_page}). The most important ones are applications of the UrQMD
model code by the CBM and PANDA collaborations \cite{CBM, PANDA} for planning of new experiments
at future FAIR facilities (GSI, Darmstadt, Germany). The code is the main event generator in the
CBM and PANDA computing frameworks. Thus it is very important to have a correct and self
consistence version of the code. Below we describe our changes of the UrQMD 1.3 package made
during our work with the code. We usually use MicroSoft Fortran Power Station for compilation
of the code. The most essential step was done when we combined all subroutines of the code into
one file and compiled it. There were too many warning messages. We have erased most of them.
The next step was when we replaced the random number generator and started to trace the code operation.
As a result, we have quite a stable and good operating code.

\section*{Changes in the file URQMD.F}
The main program is presented in the file. There the Pauli-blocking is switched off
at the end of an event generation to perform unstable particle decays, but it is not
switched on before the next event processing. Thus all calculations were performed
with no the Pauli-blocking. To improve the code operation we made the changes:

\begin{verbatim}
c optional decay of all unstable particles before final output
c DANGER: pauli-blocked decays are not performed !!!
         if(CTOption(18).eq.0) then
c no do-loop is used because npart changes in loop-structure
            i=0
            nct=0
            actcol=0
c disable Pauli-Blocker for final decays
            old_CTOption10=CTOption(10)             ! Aida
            CTOption(10)=1
c decay loop structure starts here
 40         continue
            i=i+1

c is particle unstable
            if(dectime(i).lt.1.d30) then
 41            continue
               isstable = .false.
               do 44 stidx=1,nstable
                  if (ityp(i).eq.stabvec(stidx)) then
c                     write (6,*) 'no decay of particle ',ityp(i)
                     isstable = .true.
                  endif
 44            enddo
               if (.not.isstable) then
c     perform decay
                  call scatter(i,0,0.d0,fmass(i),xdummy)
c     backtracing if decay-product is unstable itself
                  if(dectime(i).lt.1.d30) goto 41
               endif
            endif
c     check next particle
            if(i.lt.npart) goto 40
         endif ! final decay

      CTOption(10)=old_CTOption10  ! Return to the old value  ! Aida

c final output
\end{verbatim}

The variable "old\_CTOption10" is described in the declaration part of the program as

\begin{verbatim}
      integer old_CTOption10            ! Aida
\end{verbatim}

\section*{Changes in the file PROPPOT.F}

Function ERF is declared as REAL*8 ERF in various subroutines collected in the
file PROPPOT.F At the same time in file ERF.F, it is described as REAL*4, and its argument
is REAL*4 too. Due to this we had at a compilation the following diagnostics:

\begin{verbatim}
wrong data type for reference to the FUNCTION ERF from the procedure CB
the argument X (number 1) in reference to the ERF procedure from the
procedure CB is incorrect: has the wrong data type

wrong data type for reference to the FUNCTION ERF from the procedure DCB
the argument X (number 1) in reference to the procedure ERF from the
procedure DCB is incorrect: has the wrong data type

wrong data type for reference to the FUNCTION ERF from the procedure YUK
the argument X (number 1) in reference to the procedure ERF from the
procedure YUK is incorrect: has the wrong data type

wrong data type for reference to the FUNCTION ERF from the procedure DYUK
the argument X (number 1) in reference to the procedure ERF from the
procedure DYUK is incorrect: has the wrong data type
\end{verbatim}

To protect the inconsistency, we made the following replacements:

\begin{verbatim}
Original line  :     Cb = Cb0/rjk(j,k)*erf(sgw*rjk(j,k))
was replaced by:     Cb = Cb0/rjk(j,k)*erf(sngl(sgw*rjk(j,k)))  ! Aida
!                                          ^^^^^            ^

Original lines  :
      dCb = Cb0*(er0*exp(-(gw*rjk(j,k)*rjk(j,k)))*sgw*rjk(j,k)-
     +      erf(sgw*rjk(j,k)))/rjk(j,k)/rjk(j,k)

were replaced by:
      dCb = Cb0*(er0*exp(-(gw*rjk(j,k)*rjk(j,k)))*sgw*rjk(j,k)-
     +      erf(sngl(sgw*rjk(j,k))))/rjk(j,k)/rjk(j,k)          ! Aida
                ^^^^^             ^

Original lines:
      if(rjk(j,k).lt.eps) then
        Yuk = Yuk0*(er0*sgw-exp(0.25/gamYuk/gamYuk/gw)/gamYuk*
     *              (1.0-erf(0.5/gamYuk/sgw)))
      else
        Yuk = Yuk0*0.5/rjk(j,k)*exp(0.25/gamYuk/gamYuk/gw)*
     *           (exp(-(rjk(j,k)/gamYuk))*
     +            (1.0-erf(0.5/gamYuk/sgw-sgw*rjk(j,k)))-
     -            exp(rjk(j,k)/gamYuk)*
     +            (1.0-erf(0.5/gamYuk/sgw+sgw*rjk(j,k))))
      end if

were replaced by:
      if(rjk(j,k).lt.eps) then
        Yuk = Yuk0*(er0*sgw-exp(0.25/gamYuk/gamYuk/gw)/gamYuk*
     *              (1.0-erf(sngl(0.5/gamYuk/sgw))))            ! Aida
!                            ^^^^^              ^
      else
        Yuk = Yuk0*0.5/rjk(j,k)*exp(0.25/gamYuk/gamYuk/gw)*
     *           (exp(-(rjk(j,k)/gamYuk))*
     +            (1.0-erf(sngl(0.5/gamYuk/sgw-sgw*rjk(j,k))))- ! Aida
!                          ^^^^^                           ^
     -            exp(rjk(j,k)/gamYuk)*
     +            (1.0-erf(sngl(0.5/gamYuk/sgw+sgw*rjk(j,k))))) ! Aida
!                          ^^^^^                           ^
      end if

Original lines:
        dYuk = 0.5*Yuk0/rjk(j,k)*( exp(0.25/gamYuk/gamYuk/gw)*(
     *          (-(1.0/rjk(j,k))-1.0/gamYuk)*exp(-(rjk(j,k)/gamYuk))*
     *             (1.0-erf(0.5/gamYuk/sgw-sgw*rjk(j,k))) +
     *          (1.0/rjk(j,k)-1.0/gamYuk)*exp(rjk(j,k)/gamYuk)*
     *             (1.0-erf(0.5/gamYuk/sgw+sgw*rjk(j,k))) ) +
     +          sgw*er0*2.0*exp(-(gw*rjk(j,k)*rjk(j,k))) )

were replaced by:
        dYuk = 0.5*Yuk0/rjk(j,k)*( exp(0.25/gamYuk/gamYuk/gw)*(
     *          (-(1.0/rjk(j,k))-1.0/gamYuk)*exp(-(rjk(j,k)/gamYuk))*
     *             (1.0-erf(sngl(0.5/gamYuk/sgw-sgw*rjk(j,k)))) +  ! Aida
!                           ^^^^^                           ^
     *          (1.0/rjk(j,k)-1.0/gamYuk)*exp(rjk(j,k)/gamYuk)*
     *             (1.0-erf(sngl(0.5/gamYuk/sgw+sgw*rjk(j,k)))) ) +! Aida
!                           ^^^^^                           ^
     +          sgw*er0*2.0*exp(-(gw*rjk(j,k)*rjk(j,k))) )
\end{verbatim}

In addition, all "real*8 erf" have been replaced by "real*4 erf"

\section*{Changes in the file STRING.F}

We had the following diagnostic at a compilation of the file:

\begin{verbatim}
the argument MREST (number 7) in reference to procedure GETMAS
from procedure AMASS is incorrect: has the wrong data type
\end{verbatim}

The corresponding line is:
\begin{verbatim}
       call getmas(m0,w0,mindel,isoit(mindel),mmin,mmax,-1.,amass)
\end{verbatim}

The subroutine GETMAS is defined in the file DWIDTH.F:
\begin{verbatim}
      subroutine getmas(m0,g0,i,iz,mmin,mmax,mrest,m)
      integer i,iz,nrej, nrejmax
      real*8 m,m0,g0,mmin,mmax,x,x0,gg,f,g,h,pi,al,alpha,ce,mmax2
      real*8 phi,k,k0,mrest
\end{verbatim}

The argument "-1." is considered as REAL*4 at the call from the procedure AMASS.
Thus we changed the above given line in the following manner in order to
protect inconsistency.
\begin{verbatim}
!       call getmas(m0,w0,mindel,isoit(mindel),mmin,mmax,-1.,amass)  !Aida
        call getmas(m0,w0,mindel,isoit(mindel),mmin,mmax,-1.d0,amass)!Aida
!                                                           ^^
........................................................................
!        call getmas(m0,w0,ityp,iz2,mmin,mmax,-1.,amass)             !Aida
         call getmas(m0,w0,ityp,iz2,mmin,mmax,-1.d0,amass)           !Aida
!                                                ^^
\end{verbatim}

\section*{Changes in the file ANNDEC.F}

In file "tabinit.f", in "subroutine mkwtab", it is checked that the
probability of decay channel of a resonance is not zero ("bran.gt.1d-9").
If it is zero, the spline coefficients are not determined.

\begin{verbatim}
c loop over all baryons
      do 40 itp=minbar,maxbar
c get the mass of this particle
         mir=massit(itp)
c get the range of possible decay channels
         call brange (itp, cmin, cmax)
c check, if there are any decay channels
         if (cmax.gt.0) then
c loop over all decay channels
            do 41 bchan=cmin,cmax
c now get the outgoing particles 'i1' and 'i2' for the channel 'j'
c 'bran' is the mass independent branching ratio (tabulated in blockres)
c 'bflag' indicates, if 'i1', 'i2' or both are broad
               call b3type (itp,bchan,bran,i1,i2,i3,i4)
c check, if decay is allowed

               smass=mminit(i2)
               if(i3.ne.0) smass=smass+mminit(i3)
               if(i4.ne.0) smass=smass+mminit(i4)

               if (bran.gt.1d-9.and.mir.gt.mminit(i1)+smass) then
c loop over all x-values
                  do 42 i=1,widnsp
c store the values
                     pbtaby(i,1,itp,bchan)=
     .                    fbrancx (bchan,itp,isoit(itp),tabx(i),
     .                    bran,i1,i2,i3,i4)
 42               continue
c calculate the second derivate and store it in 'pbtaby(,2,,)'
                  call spline (tabx(1),pbtaby(1,1,itp,bchan),widnsp,
     .                         abl0,abln,pbtaby(1,2,itp,bchan))
               end if
 41         continue
         end if
 40   continue
      write (6,*) '(3/7) ready.'
\end{verbatim}

At the same time, in the file anndec.f, in subroutine anndex,
it is not checked that the probability is zero.

\begin{verbatim}
C   one ingoing particle --> two,three,four outgoing particles
C
c... decays
         do 3 i=0,maxbr
            if(isoit(btype(1,i))+isoit(btype(2,i))+isoit(btype(3,i))+
     &         isoit(btype(4,i)).lt.iabs(iz1).or.
     &           m1.lt.mminit(btype(1,i))+mminit(btype(2,i))
     &                +mminit(btype(3,i))+mminit(btype(4,i)) )then
               prob(i)=0.d0
            else
               prob(i)=fbrancx(i,iabs(i1),iz1,m1,branch(i,iabs(i1)),
     &              btype(1,i),btype(2,i),btype(3,i),btype(4,i))
            endif
 3       continue
\end{verbatim}

\noindent Thus a call of "fbrancx" was performed for a channal which was
not described for the spline interpolation.

To improve the situation, we have added many lines in the subroutine anndex.
\begin{verbatim}
C   one ingoing particle --> two,three,four outgoing particles
C
c... decays

 do 3 i=0,maxbr
   if((minbar.le.iabs(i1)).and.(iabs(i1).le.maxbar)) then        ! Uzhi
     call b3type (i1,i,bran_uz,i1_uz,i2_uz,i3_uz,i4_uz)          ! Uzhi
     if(bran_uz.le.1.d-9)                    then     ! Uzhi see mkwtab
         prob(i)=0.d0                                            ! Uzhi
     else                                                        ! Uzhi
       if(isoit(btype(1,i))+isoit(btype(2,i))+isoit(btype(3,i))+ ! Uzhi
&         isoit(btype(4,i)).lt.iabs(iz1).or.                     ! Uzhi
&           m1.lt.mminit(btype(1,i))+mminit(btype(2,i))          ! Uzhi
&                +mminit(btype(3,i))+mminit(btype(4,i)) )then    ! Uzhi
          prob(i)=0.d0                                           ! Uzhi
       else                                                      ! Uzhi
          prob(i)=fbrancx(i,iabs(i1),iz1,m1,branch(i,iabs(i1)),  ! Uzhi
&              btype(1,i),btype(2,i),btype(3,i),btype(4,i))      ! Uzhi
       endif                                                     ! Uzhi
     endif                                                       ! Uzhi
   else                                           ! For mesons   ! Uzhi

       if(isoit(btype(1,i))+isoit(btype(2,i))+isoit(btype(3,i))+
&         isoit(btype(4,i)).lt.iabs(iz1).or.
&           m1.lt.mminit(btype(1,i))+mminit(btype(2,i))
&                +mminit(btype(3,i))+mminit(btype(4,i)) )then
          prob(i)=0.d0
       else
          prob(i)=fbrancx(i,iabs(i1),iz1,m1,branch(i,iabs(i1)),
&              btype(1,i),btype(2,i),btype(3,i),btype(4,i))
       endif
      endif                                                      ! Uzhi
3 continue
\end{verbatim}

In addition, we have added description of the variables
bran\_uz, i1\_uz, i2\_uz, i3\_uz, i4\_uz in the declaration part of the subroutine
\begin{verbatim}

      real*8 bran_uz                             ! Uzhi
      integer i1_uz,i2_uz,i3_uz,i4_uz            ! Uzhi
\end{verbatim}

\section*{Change in the file BLOCKRES.F}
We had a problem with decay of $\Delta (1950)$. Thus we have changed
probabilities of the decay channels looking at UrQMD 1.2 code.
\begin{verbatim}
c delta resonances
 a  6d-3, 1.0, .00, .00, .00, .00, .00, .00, .00, .00, .00, .00, !1232
 b    0., .10, .00, .00, .00, .00, .65, .25, .00, .00, .00, .00, !1600
 c  4d-4, .15, .00, .00, .05, .00, .65, .15, .00, .00, .00, .00, !1620
 d  2d-3, .20, .00, .00, .25, .00, .55, .00, .00, .00, .00, .00, !1700
 e    0., .25, .00, .00, .25, .00, .25, .25, .00, .00, .00, .00, !1900
 f  3d-4, .18, .00, .00, .80, .00, .02, .00, .00, .00, .00, .00, !1905
 g    0., .30, .00, .00, .10, .00, .35, .25, .00, .00, .00, .00, !1910
 h    0., .27, .00, .00, .00, .00, .40, .30, .00, .03, .00, .00, !1920
 i    0., .22, .00, .00, .05, .00, .40, .30, .00, .03, .00, .00, !1930
 j 15d-3, .38, .00, .00, .00, .00, .34, .24, .00, .00, .00, .00/ !1950 !Uzhi
c j 15d-3, .38, .00, .00, .00, .00, .34, .24, .00, .00, .00, .04/ !1950!Uzhi
\end{verbatim}

\section*{Change in the file INIT.F}
In order to trace the code operation, we changed the value of parameter ("nnuc=11"),
and put it to "1".
\begin{verbatim}
    parameter (nnuc=1)    ! 1)                           ! Uzhi
\end{verbatim}

In the corresponding subroutine, the coordinates and momentum components of nuclear
nucleons are sampled for 1, 11, 22, 33 and so on events. For other events with
intermediate numbers, the quantities are obtained by randomly rotation of the sampled
coordinates and momenta. With the new value of "nnuc" the quantities are sampled for each
event.

\section*{Saving of local variables}
Many fortran users believe that the variables initiated in a program unit with
the help of DATA operators are stored during full time of program work. It is true only
for a first usage of the variables at some computers. After that the values are changed
in an unpredictable manner. Thus FORTRAN standard requires to save the variables with
the help of SAVE operator.

There are a lot of such variables in the UrQMD code. We were trying to save most of them.

\begin{verbatim}
angdis.f:      SAVE symlog                                       ! Aida
angdis.f:      SAVE msi, cmsi,gsi, mom,cmom,gom, mpi,cmpi,gpi,m  ! Aida
ityp2pdg.f:    save idtab                                        ! Aida
ityp2pdg.f:    save baryon_names                                 ! Aida
ityp2pdg.f:    save meson_names                                  ! Aida
make22.f:      save ar                                           ! Aida
make22.f:      save rr                                           ! Aida
make22.f:      save rr                                           ! Aida
make22.f:      save in                                           ! Aida
string.f:      save mixang                                       ! Aida
string.f:      save AMq                                          ! Aida
\end{verbatim}

\end{document}